\documentstyle[letter,twocolumn,epsf]{jpsj}
\newcommand{\bm}[1]{\mbox{\boldmath$#1$}}
 \makeatletter
 \def\lsim{\mathrel{\mathpalette\gl@align<}}
 \def\gsim{\mathrel{\mathpalette\gl@align>}}
 \def\gl@align#1#2{\lower.6ex\vbox{\baselineskip\z@skip\lineskip\z@
 \ialign{$\m@th#1\hfil##\hfil$\crcr#2\crcr\sim\crcr}}}
 \makeatother
\title{
High Temperature Expansion for the SU($n$) Heisenberg Model\\
in One Dimension
}
\author{Noboru {\sc Fukushima}${}^{1,2}$\footnote{e-mail address:
fuku@mpipks-dresden.mpg.de} and Yoshio \sc{Kuramoto}${}^2$}
\inst{
${}^1$ Max-Planck-Institut f\"ur Physik komplexer Systeme,
N\"othnitzer Stra{\ss}e 38, D-01187 Dresden, Germany \\
${}^2$ Department of Physics, Tohoku University, Sendai 980-8578, Japan
}

\recdate{ }
\abst{
Thermodynamic properties of the SU($n$) Heisenberg model
in one dimension is studied by means of high-temperature expansion
for arbitrary $n$.
The specific heat
up to $O[(\beta J)^{23}]$
and
the correlation function
up to $O[(\beta J)^{18}]$ are derived with $\beta J$ being the
antiferromagnetic exchange in units of temperature.
It is found for $n>2$ that the specific heat shows a shoulder in the high-temperature side of a peak.
The origin of this structure is clarified by deriving the temperature dependence of the correlation function.
With decreasing temperature, the short-range correlation with two-site periodicity develops first, and then another correlation with $n$-site periodicity at lower temperature.
This behavior is in contrast to that of the inverse square interaction model, where
the specific heat shows a single peak according to the exact solution.
Our algorithm has an advantage that neither computational time nor memory depends on the
multiplicity $n$ per site; the series coefficients
are obtained as explicit functions of $n$.
}
\kword{
high temperature series expansion, orbital degeneracy,
Heisenberg model, SU($n$) symmetry, multipolar interaction, fluctuation,
exchange interaction model
}

\begin{document}
    \sloppy
    \maketitle

Recently, orbitally degenerate systems attract much attention
as strongly-correlated electron systems.
The orbital degeneracy is regarded as an internal degrees of freedom in addition to the spin degeneracy.  The total number $n$ of internal degrees of freedom thus varies
depending on the system.
If the number $n^2-1$ of competing fluctuations is large,
the transition temperature of the orbital ordering is
significantly suppressed from the
the mean-field value
even in three dimensions\cite{KF}.
Thus reliable theoretical scheme
beyond the mean-field theory should be necessary
to investigate orbitally degenerate systems.

As the simplest model to describe the orbitally degenerate systems,
the SU($n$) Heisenberg model has been investigated.
In one dimension, it can be solved exactly by the Bethe ansatz\cite{Sutherland}.
However, some interesting quantities like the correlation function are difficult to be derived by the Bethe ansatz.
Some critical exponents have been calculated with use of the non-abelian
bosonization\cite{Affleck}
and the conformal field theory\cite{Kawakami}.
Numerical methods have also been used to study the system.
For $n=4$, the spin-spin correlation function has been calculated
using the density matrix renormalization group\cite{DMRG}
and the quantum Monte Carlo method\cite{QMC}.  The latter has also derived
temperature dependence of the correlation lengths and the entropy.
Although these numerical approaches are powerful, they are not always convenient to investigate systems with large $n$, or in higher dimensions.

As an alternative approach which is applicable to any dimension,
the high-temperature expansion has provided significant information on variety of models \cite{Domb3}.
We have recently developed an efficient algorithm to carry out
the high-temperature expansion for systems in which the Hamiltonian is written
in terms of exchange operators.
With use of this new algorithm,  we report in this paper a peculiar temperature dependence of the spatial correlation, which becomes conspicuous as
the number $n$ of internal degrees of freedom increases.
Namely, with decreasing temperature, the short-range correlation with two-site periodicity develops first, and then another correlation with $n$-site periodicity at lower temperature.
This leads to a corresponding structure in the specific heat.

In our algorithm, neither computational time nor
memory depends on the multiplicity $n$ per site;
$n$ is just a parameter --- the series coefficients
are obtained as explicit functions of $n$.
Details of the algorithm as well as application to higher dimensional models will be reported  separately.

We consider a one-dimensional lattice of $N$ sites with a periodic boundary condition.
Let each site take one of the $n$ colors,
and denote them as $|\alpha\rangle$ with $\alpha = 1, 2, \cdots, n$.
We define
$X^{\alpha\beta}:= |\alpha\rangle\langle\beta|$
and an exchange
(or permutation) operator
\begin{equation}
 P_{i,j}:= \sum_{\alpha=1}^n  \sum_{\beta=1}^n
X_i^{\alpha\beta} X_j^{\beta\alpha}.
\end{equation}
Colors of sites $i$ and $j$ are exchanged when $P_{i,j}$ is applied.
The Hamiltonian is given by
\begin{equation}
 {\cal H}:=  J \sum_{i=1}^{N-1} P_{i,i+1} + J P_{1,N}.
\label{hamil}
\end{equation}
Then, the Hamiltonian has the SU($n$) symmetry.
We investigate the antiferro-interaction model
with $J>0$.
When $n=2$, this model is equivalent to the ordinary Heisenberg model.
The SU(4) Heisenberg model
is related to orbital- and spin-degenerate
systems. The local four states
are denoted as
$|+)|+]$, $|+)|-]$, $|-)|+]$, $|-)|-]$,
where $|\pm)$ and $|\pm]$ stand for a orbital state and a spin
state respectively. The pseudo-spin operators are defined
as $t^z|\pm)=\pm\frac12|\pm)$, $t^\pm|\mp)=|\pm)$,
$s^z|\pm]=\pm\frac12|\pm]$, $s^\pm|\mp]=|\pm]$, and then
the exchange operator for $n=4$ is rewritten as
\begin{eqnarray}
P_{i,j} &=& \bm{t}_i\cdot\bm{t}_{j} + \bm{s}_i\cdot \bm{s}_{j}
  +4 (\bm{t}_i\cdot\bm{t}_{j})  (\bm{s}_i\cdot \bm{s}_{j})
 +\frac{1}{4},
\end{eqnarray}
or,
\begin{eqnarray}
4 P_{i,j}
&=&\bm{\tau}_i\cdot\bm{\tau}_{j}
+ \bm{\sigma}_i\cdot \bm{\sigma}_{j}
  + (\bm{\tau}_i\cdot\bm{\tau}_{j})  (\bm{\sigma}_i\cdot \bm{\sigma}_{j})
 +1
\label{paulirep}
\end{eqnarray}
using the Pauli matrices
$\bm{\tau}=2\bm{t}$ and $\bm{\sigma}=2\bm{s}$.

The high temperature expansion
is performed by expanding
the Boltzmann factor $e^{-\beta {\cal H}}$
in $\beta$.
To investigate the specific heat,
we calculate $\langle P_{i,i+1} \rangle$
because the internal energy is obtained by
$\langle {\cal H} \rangle=N J \langle P_{i,i+1} \rangle$.
We also investigate correlation functions.
For $n=4$, fifteen components in eq.(\ref{paulirep})
contribute equally, and thus
$
\langle \tau_i^\alpha \tau_j^\alpha\rangle
=
\langle \sigma_i^\beta \sigma_j^\beta\rangle
=\langle \tau_i^\gamma \sigma_i^\delta
         \tau_j^\gamma \sigma_j^\delta \rangle
=
\langle 4 P_{i,j} -1 \rangle /15 $.
For general $n$,
there is a relation,
\begin{equation}
\langle X_i^{\alpha\beta} X_j^{\beta\alpha} \rangle=
\frac{1}{n^2-1} \left( \langle  P_{i,j} \rangle -\frac{1}{n} \right),
\end{equation}
for $i\neq j, \alpha \neq \beta$.

Although the concept of the high temperature expansion is quite simple,
it needs various
devices to calculate terms of high orders.
We mainly use the finite cluster method\cite{Domb3, Gelfand}.
In this method,
series coefficients in the thermodynamic limit are obtained exactly
by weighted-summation of those calculated in finite-size clusters.
The cluster of size $\ell$ is defined by
\begin{equation}
{\cal H}_\ell := J \sum_{i=1}^{\ell-1} P_{i,i+1}.
\end{equation}
To obtain the series expansion of
$\langle P_{i,j} \rangle$ up to $O[(\beta J)^M]$,
we need to calculate ${\rm Tr}[({\cal H}_\ell)^m]$ and
${\rm Tr}[P_{i,j}({\cal H}_\ell)^m]$ for $0\le m\le M$.
In calculating the trace, we adopt a method to use
combinatorics\cite{loop, chen}.
Then, coefficients are obtained as explicit functions of $n$.
Recently a related expansion for the specific heat and the uniform susceptibility
was reported for the Heisenberg model ($n=2$)\cite{uhrig}.  We have checked that the series obtained by our method reproduces the previous result.

In using our method, we have a trick to store data
and reduce computational time and memory.
Further, we use another trick to reduce
the required maximum system size $\ell_{\rm max}$.
If only the finite cluster method is used,
the series for $\langle P_{i,i+1} \rangle$ up to $O[(\beta J)^M]$
requires
$\ell_{\rm max}=M/2+1$ ($M$: {\rm even})
or $\ell_{\rm max}=(M+1)/2+1$ ($M$: {\rm odd}),
and the Fourier transform of $\langle P_{i,j} \rangle$
requires $\ell_{\rm max}=M+1$.
However, we reduce $\ell_{\rm max}$
by formulating a method complementary to
the finite cluster method.
Namely, we can calculate $\langle P_{i,i+1} \rangle$
up to $O[(\beta J)^{22}]$ with $\ell_{\rm max}=11$.
The Fourier transform of the correlation function
$\langle P_{i,j} \rangle$ up to $O[(\beta J)^{18}]$
is calculated with $\ell_{\rm max}=12$.
As for the specific-heat in the $n\rightarrow \infty$ limit,
coefficients of $(\beta J)^{m}$ with odd $m$ are equal to zero,
and non-zero coefficients are calculated up to $O[(\beta J)^{28}]$.

We perform the Pad\'{e} approximation (PA) and
the first-order inhomogeneous differential
approximation\cite{Domb13}(FOIDA) to extrapolate the series.
In general, the FOIDA shows better convergence than the PA.
However, changing the expansion variable can
improve the convergence of the PA.
We change the variable from $\beta J$ to $\tanh (c \beta J)$,
where $c$ is an adjustable parameter to improve convergence\cite{tanh}.
For small $c$, the results do not differ much from those in $\beta J$.
For large $c$, the convergence becomes better.
However the $\tanh$ variable saturates for $\beta J$ much larger than $1/c$,
and the approximant tends to saturate accordingly.
Therefore, we take the smallest possible $c$ as long as the convergence remains good.
We refer to this extrapolation as the PAT hereafter.

\begin{figure}[h]
\begin{center}
\epsfxsize = 8.5cm \epsfbox{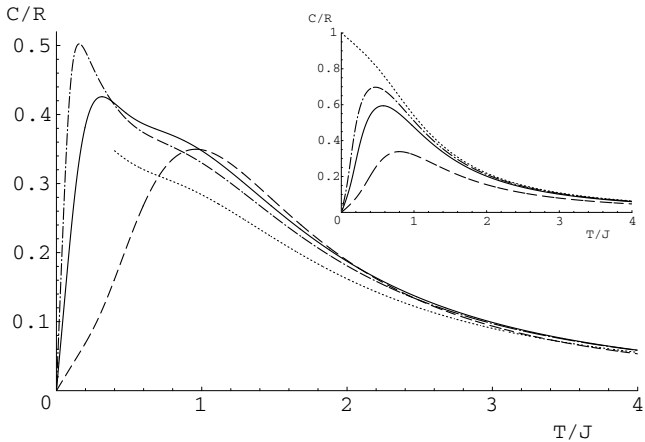}
\caption{
Specific heat for $n=2$ (dashed line), $n=4$ (solid line),
$n=6$ (dash-dotted line) and $n=\infty$ (dotted line),
in units of the gas constant $R=8.31 {\rm J/(mol \cdot K)}$.
The $n=\infty$ data are plotted only for $T > 0.4 J$.
The inset shows the exact solution of the inverse-square interaction model
\cite{kurakato, comment:r2}.
\label{fig:hinetu}}
\end{center}
\end{figure}

To investigate the specific heat $C(T)$,
we first analyze the series for the entropy $S(T)$ and then
calculate $C=T {\rm d} S/{\rm d} T$
to use the known fact $S(T\rightarrow\infty)=\log n$.
For the PAT, we extrapolate $S/T$ rather than $S$
because the exact solution for the ground state\cite{Sutherland}
behaves as $S \propto T$ at $T \rightarrow 0$.
When $n=4$, our result for the entropy
shows a good agreement to
that of the quantum Monte Carlo method\cite{QMC}.
Although the behavior of the entropy seems simple at a sight,
the specific heat clearly shows a structure
composed of a peak at
low temperature and a shoulder at
higher temperature.
Figure \ref{fig:hinetu} shows the results of
the specific heat for various values of $n$ using the PAT.
At $T \gsim 0.4J$,
the results of the PAT and the FOIDA coincide within the width of the line,
and thus the estimated error is  smaller than the width.
The high temperature limit is given by $C \rightarrow (1-n^{-2})\beta^2 J^2$.
Note that the existence of the shoulder structure
is clearly seen,
which accompanies changes of the sign of the curvature.
In this range the expansion converges well.
The position of the shoulder seems to be independent of $n$,
and the structure remains in the limit $n\rightarrow \infty$.
We shall discuss the origin of this structure later in this paper.

As $n$ increases, the peak
in the specific heat
becomes sharper and shifts to lower temperature.
Such a rapid change of a function at very low temperature
is difficult to extrapolate.
Hence the error becomes larger with increasing $n$.
Particularly, the FOIDA
becomes unstable at $T\lsim 0.4J$;
the results of the FOIDA are scattered, whereas
some results agree with that of the PAT.
This is because we do not use any information at $T\rightarrow0$
for the FOIDA.
As for the PAT,
we estimate that the error at $T \simeq 0.2J$ is about 5\% when $n=6$,
and smaller than it when $n<6$.
The specific heat at $T\rightarrow 0$ using the PAT
is
by 5\% $\sim$ 15\% larger than the exact value
$\frac{1}{6}n(n-1) T$,
which is derived from the exact solution\cite{Sutherland}.

We remark that the behavior of the specific heat is different from
that of the $1/r^{2}$-model\cite{halshas} defined by
$$
{\cal H}_{1/r^{2}}:=J\sum_{i<j}[(N/\pi)\sin\{\pi(i-j)/N\} ]^{-2}P_{i,j}.
$$
The inset of Fig.\ref{fig:hinetu} shows
the exact result \cite{kurakato, comment:r2} for the $1/r^{2}$-model,
which does not show the peak-shoulder structure
in contrast with the nearest-neighbor model.
Furthermore the $T\rightarrow0$ limit of the specific heat
is given by $\frac{1}{3}(n-1) T$,
which is $2/n$ of the nearest-neighbor case.
We interpret this difference as coming from the
different $n$-dependence of the spinon velocity.
\cite{Sutherland, yamamoto}
Namely, the spinon velocity in the nearest neighbor model decreases as $1/n$ for large $n$, while the velocity stays constant in the $1/r^{2}$-model.
Such difference is understood intuitively by considering the supersymmetric {\it t-J} model \cite{kurayoko} and eliminating the holes.
The spinon velocity is related to the Fermi velocity in the non-interacting counterpart and reflects its spectrum.
The $1/r^{2}$-model has a cusp in the bottom of the band, which leads to the finite velocity even for vanishing Fermi energy.
While the cosine band of the nearest-neighbor model gives $1/n$-dependence for the Fermi velocity.

\begin{figure}[h]
\begin{center}
\epsfxsize = 8cm \epsfbox{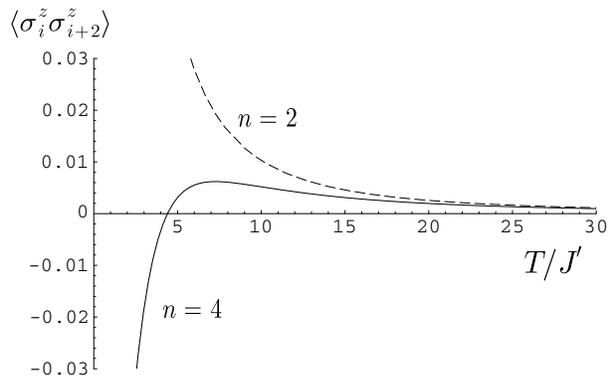}
\caption{
Temperature dependence of
the next-nearest-neighbor correlation
$\langle \sigma^z_i \sigma^z_{i+2} \rangle$ for $n=2, 4$.
The estimated error is within the width of the line.
The unit of temperature is
given by $J'=J/n$.
\label{s1s3}}
\end{center}
\end{figure}

In order to see the temperature-dependent nature of the spatial correlation,
we show in
Fig.\ref{s1s3} the next-nearest-neighbor correlation
$\langle \sigma^z_i \sigma^z_{i+2} \rangle$ for $n=2, 4$.
Here we have introduced $J':=J/n$ so as to
match the correlation function in the high-temperature limit.
We compare systems with $n=2$
$$
J'\sum \bm{\sigma}_i\cdot \bm{\sigma}_{j},$$
and $n=4$
$$
J'\sum \left\{\bm{\tau}_i\cdot\bm{\tau}_{j}
+ \bm{\sigma}_i\cdot \bm{\sigma}_{j}
  + (\bm{\tau}_i\cdot\bm{\tau}_{j})  (\bm{\sigma}_i\cdot
\bm{\sigma}_{j})\right\}.
$$
The lowest order of the series of
$\langle \sigma^z_i \sigma^z_{i+x} \rangle$
is $O[(\beta J)^x]$ and its sign is given by $(-1)^x$.
Then we have
$\langle \sigma^z_i \sigma^z_{i+2} \rangle>0$ at high temperature.
As temperature decreases,
$\langle \sigma^z_i \sigma^z_{i+2} \rangle$ first grows larger,
then starts to decrease and turn to
negative when $n=4$.
The reason
is the following:
Each of 15 components
tries to align itself antiparallel.
However, this cannot be
attained simultaneously ---
e.g. the N\'{e}el states of both $\tau^z$ and $\sigma^z$
makes ferromagnetically aligned $\tau^z\sigma^z$.
Therefore, as each antiferromagnetic correlation becomes larger,
each short-range order disturbs each other,
and thus the correlation turns to a longer period
at low temperature.

\begin{figure}[h]
\begin{center}
\epsfxsize = 7cm \epsfbox{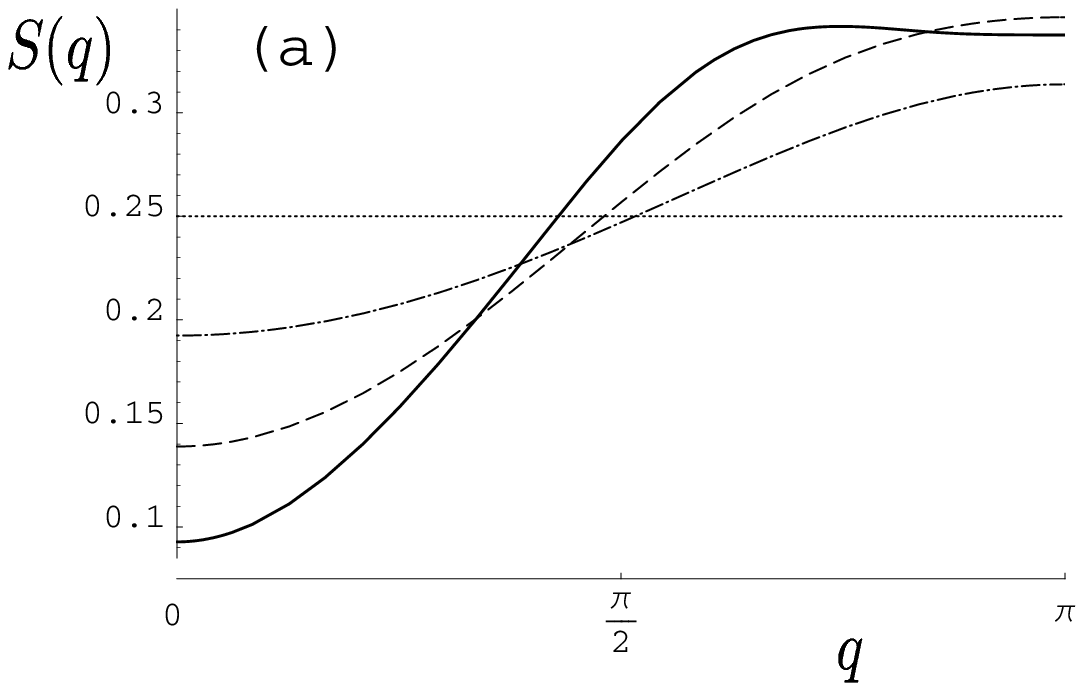}
\epsfxsize = 7cm \epsfbox{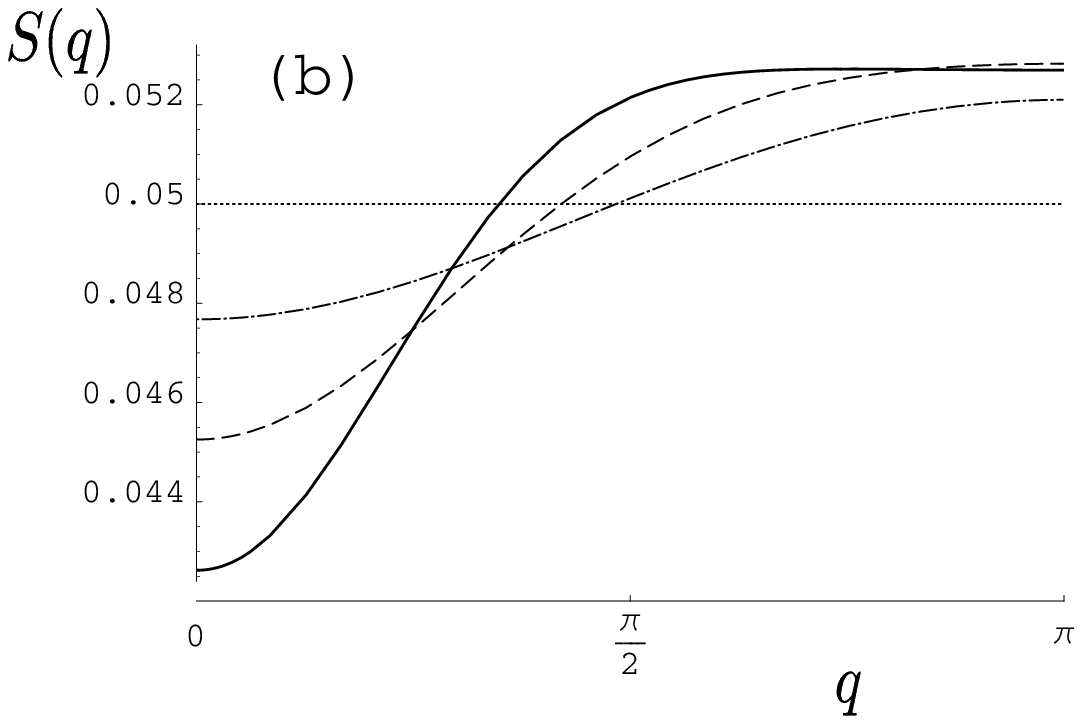}
\caption{
Fourier Transform $S(q)$ of the correlation function
for (a) $n=4$ and (b) $n=20$
at temperature $T=\infty$ (dotted line), $T=2 J$ (dash-dotted line),
$T=0.8 J$ (dashed line) and $T=0.4 J$ (solid line).
The estimated error is within the width of the line.
At $T=0.4 J$, the maximum is located at $q\ne\pi$ in both
(a) and (b).
\label{fig:sq}}
\end{center}
\end{figure}

Such a change of correlation at low temperature
also appears in
the Fourier transform of the correlation function
for $n>2$.
We define
$S(q):= \sum_{x}
\langle X_j^{\alpha\beta} X_{j+x}^{\beta\alpha} \rangle e^{-iqx}
\;(\alpha\neq\beta)$, which does not depend on $\alpha$ or $\beta$
because of the SU($n$) symmetry.
Thus it is
reduced to $S(q)= \sum_{x}
\langle s_j^z s_{j+x}^z \rangle e^{-iqx}$ when $n=4$.
A naive extrapolation of
the series for $S(q)$ shows bad convergence at several
values of $q$.
This is because information at distance $x$
is lacking when $\cos(x q)\simeq 0$.
To avoid this difficulty we extrapolate the series of a complex function
$2 \sum_{x>0}
\langle X_j^{\alpha\beta} X_{j+x}^{\beta\alpha} \rangle e^{-iqx}$,
and take the real part of the extrapolated function.
This method makes use of its imaginary part
as well. As a result
the convergence becomes better than the naive extrapolation.
Figure \ref{fig:sq} shows the result
at several different temperatures for $n=4, 20$.
The results of the PAT and the FOIDA coincide within the width of the line.
In the high temperature limit,
only the on-site contribution
$\langle X_j^{\alpha\beta} X_{j}^{\beta\alpha} \rangle $
remains and
$S(q)$ becomes a constant $1/n$.
As temperature decreases, $S(\pi)$ increases first, and
reaches the maximum at $T_{\pi{\rm max}}\simeq 0.8J$.
With further decrease of temperature, $S(\pi)$ starts decreasing.
Finally
the maximum of $S(q)$ starts moving
from $q=\pi$
at $T_{\rm mv}\simeq 0.5 J$.
Comparing the different $n$ results at the same temperature,
the peak of $S(q)$ around $q=\pi$ becomes broader as $n$ increases,
especially at lower temperature.
However, $T_{\pi{\rm max}}$ and $T_{\rm mv}$ seem to be
independent of $n$, even in the $n\rightarrow \infty$ limit.

Other numerical studies \cite{DMRG, QMC} for $n=4$
show that $S(q)$ at zero temperature
has cusps at $q=0, \pi/2, \pi$
and  has the maximum at $q=\pi/2$.
For general $n$,
the dominant correlation in the ground state
\cite{Sutherland,Affleck,Kawakami} is that of $q=2\pi/n$,
and the maximum of $S(q)$ is expected to reach $q=2\pi/n$
in the $T\rightarrow 0$ limit.
The Fourier transform leading to the cusp type behavior requires data of spatial correlations for large distance.
In Fig.\ref{fig:sq} the temperatures are still too high to observe this behavior.

A plausible characteristic temperature of the long-range correlation
is given by the mean-field approximation.
Note that the transition temperature in this approximation is given by
$T_{\rm N}^{\rm MF}=2J/n$,
which decreases with increasing $n$.
Therefore, if we scale energies by $T_{\rm N}^{\rm MF}$,
$S(q)$ starts to shift
toward the ground-state correlation
at higher temperature
as $n$ increases.
Namely $T_{\rm mv}/T_{\rm N}^{\rm MF}$ increases.

From the results of the specific heat and the correlation function,
we consider the reason why the structure
appears in the specific heat.
Let us
imagine cooling down the system gradually
from the high temperature limit.
In the case of the nearest-neighbor interaction,
short-range correlation with two-site periodicity
develops first at high temperature.
It makes energy gain and
the specific heat grows as temperature decreases.
However, since the dominant correlation in the ground state
is not the $q=\pi$ component,
its energy gain saturates.
This saturation should cause the shoulder of the specific heat.
Then, there is another energy gain
when the
system begins to achieve
the long-range correlation with $q=2\pi/n$.
It should cause the peak at the temperature.

One may expect that the shift of the peak position of the specific heat
is mainly due to the shift of $T_{\rm N}^{\rm MF}=2J/n$ to
lower temperature with increasing $n$.
However,
the peak position
decreases more rapidly
than the $1/n$ behavior.
One of the reasons should be that
 $T_{\rm N}^{\rm MF}$ is not for the $q=2\pi/n$ component
but for the $q=\pi$ component.
Strong quantum effects are required to realize
the $q=2\pi/n$ correlation.
Accordingly a temperature lower than $T_{\rm N}^{\rm MF}$ gives a peak in the specific heat.

For comparison, let us
consider the $1/r^{2}$-model.
There are interactions even in the intra-sublattice of the N\'{e}el state.
These make the short-range two-site periodic correlation
less favorable\cite{comment} even at high temperature.
Further,
interactions making different colors {\it attractive}, and
the same colors {\it repulsive} are long-ranged.
Then the $n$-site periodic ($q=2\pi/n$) correlation is
favorable at low temperature.
As a result, the same
correlation as
that of the ground state should develop
gradually from high temperature.
The specific heat therefore shows a single peak.

In conclusion, we find by large orders of  high-temperature expansion
a peak-shoulder structure in the specific heat
which results from different behaviors
of spatial correlations at high and low temperatures.
Namely the short-range correlation at high temperature
has a two-site periodicity,
while the ground-state correlation has
the $n$-site one.
We will discuss applications of the high-temperature expansion
to higher dimensional systems in the near future.

\acknowledgement
We are grateful to R.~R.~P.~Singh for fruitful discussions
about series expansion methods.
We also thank to M. Koga and M. Arikawa for discussions.
N.~F. would like to thank
the Japan Society for the Promotion
of Science and the Visitors Program of the MPI-PKS for the support.


\begin{thebibliography}{130}
\bibitem{comment:r2}
The $n\rightarrow\infty$ limit is not uniformly
convergent. Namely, although $C(T=0)=0$ for any $n$,
all the $C(T>0)$ values converge to the dotted-line shown
in the inset of Fig.\ref{fig:hinetu}.
\bibitem{comment}
In the  high temperature limit, the correlation function
is
$0>\langle X_j^{\alpha\beta} X_{j+r}^{\beta\alpha} \rangle \propto r^{-2}$
and its Fourier transform $S(q)$ has the maximum at $q=\pi$.
However, the peak around $q=\pi$ is much broader than that of
the nearest-neighbor model.
\end{thebibliography}
\end{document}